\documentclass{article}

\usepackage{comment}
\usepackage[affil-it]{authblk}
\usepackage[utf8]{inputenc}
\usepackage{amsthm,amsmath,amssymb,amsfonts,latexsym,graphicx,enumerate,setspace,verbatim,tocloft,rotating}
\theoremstyle{plain} 
\newtheorem{theorem}{Theorem}

\newtheorem{corollary}[theorem]{Corollary}

\theoremstyle{definition}  

\theoremstyle{remark}  

\title{An Invariant Between Hyperbolic Surfaces and Lattice Spin Models}
\author{William Chuang\thanks{Electronic address: \texttt{williamchuang@math.arizona.edu}}}
\affil{Department of Mathematics,\\ University of Arizona, Tucson, USA}

\date{\today}

\begin{document}

\maketitle

\begin{abstract}
	In this succinct note, it is showed that a partition function of equivalent classes of hyperbolic surfaces can be connected to an Ising model located on the boundary of the Poincar\'{e} disc, as hinted by Poincar\'{e}'s Uniformization theorem and Patterson-Sullivan's Theorem.
\end{abstract}

\section{Introduction}

There are several methods for researching quantum gravity. The path-integral formalism is the method used in this brief note. If we are able to determine all of a particle's classical pathways, we may then develop a quantum theory of the particle's motion using the calculus of variation, also known as the path integral technique. Similar to this, in order to quantize a spacetime of a universe, one must be aware of all potential outcomes, or all classical spacetime solutions (in this note they are equivalent classes of hyperbolic surfaces). To begin with, this note is built on hyperbolic spaces, which can be in two-dimension or in some higher dimensions\cite{MR725161,beardon2012geometry,MR2985759,MR945013,MR1138441,MR2402415,MR4221225,MR1893917,MR2040008}.

In order to keep things simple, the author utilized 2d hyperbolic space, which is isomorphic to 2+1 Anti-de Sitter (Ads) space, as the main example. Please keep in mind that although propagating gravitational degrees of freedom is not possible in 2+1 dimensions, this dimension nevertheless functions as a conceptual model for the topology and geometry of the universe\cite{carlipMR1637718,carlip2005quantum}.

This brief note can also be applied to some higher dimensions because the correspondence theorem\cite{sullivan1979density}, on which the primary conjecture is founded, has also been generalized to some higher dimensions\cite{bishop1997hausdorff}. For $n$-dimension hyperbolic manifold\cite{MR725161,beardon2012geometry,MR2985759,MR945013,MR1138441,MR2402415,MR4221225,MR1893917,MR2040008}, one can replace $\mathbb{H}^2\setminus\Gamma$ to the higher dimensional algebraic curve $\mathbb{H}^n\setminus\Gamma$ and the main theorem of this note follows. This note focuses on the lower dimensions since these cases are considered as the base case for constructing the result of the main theorem in higher dimensions.

The critical exponent of the Poincar\'{e} series of the Schottky group $\delta$ is equal to the Hausdorff dimension $H_{L(\Gamma)}$ of the limit set of a convex cocompact Schottky group, according to the 2d version of this correspondence theorem. Beardon \cite{beardon1968exponent,beardon1971inequalities} established one direction first, followed by Patterson and Sullivan who finished the other\cite{patterson1976limit,sullivan1982discrete,sullivan1984entropy}.

The other main foundation is Poincar\'{e} Uniformization Theorem. It states that if there exists a polygon $F$ with a side pairing $\Phi$, then the group generated by side pairing must be discrete and the space $S$ can be tessellated by using all elements in the discrete group $G$. Then the quotient space $S\setminus G$ (or in some books it is denoted as $S/G$ which is also a Riemann surface and an algebraic curve and it means the space $S$ can be tessellated by using the fundamental domain with all elements in the group $G$). Furthermore, this space $S\setminus G$ will admit a Riemann metric with constant curvature and so $S$ can be a 2d-hyperbolic space, 2d-Euclidean space, or a sphere.

Conversely, a compact Riemann surface (algebraic curve) is a quotient space $S\setminus G$ where $S$ can be a 2d-hyperbolic space, 2d-Euclidean space, or a sphere and $G$ is a group of Mobi\"{u}s maps on the complex plane. Furthermore, in this direction, a covering  will be induced, and we can have $\theta: \Gamma\setminus D(G)\to \Gamma \setminus \mathbb{D}$ is a homeomorphism, if and only if the fundamental domain $D(G)$ is locally finite, and this $D(G)$ is a hyperbolic polygon, or it is also called the fundamental polygon of the Riemann surface. Some more details and examples can be found in Chapter 9 of Beardon's book\cite{beardon2012geometry}.

Let $\mathbb{D}$ be a Poincar\'{e} model of hyperbolic space, $\partial\mathbb{D}$ be the boundary of the hyperbolic space, $\Gamma$ be a geometrically finite non-elementary convex co-compact Schottky group, $L(\Gamma)$ be the limit set of $\Gamma$, and $\Gamma\setminus\mathbb{D}$ be the hyperbolic manifold. Furthermore, since $\Gamma$ is convex co-compact, $\Gamma\setminus\mathbb{D}$ is conformally compact\cite{borthwick2007spectral}.

Then, there are quite a few other correspondences which can also shed some light on this boundary and interior bulk correspondence. This critical exponent $\delta$ is related to the the first resonance $\lambda_0$ of the Laplacian acting on the hyperbolic surface: $\delta(1-\delta)=\lambda_0$\cite{patterson1987lectures,patterson1988lattice}. The corollary of this result also mimic the importance of the $s=1$ line on the complex plane for the fundamental theorem used in the the proof of Prime Number Theorem\cite{jameson2003prime} that this can be used to prove Prime Geodesic Theorem for cases of modular groups, since for modular groups, $\delta=1$. Then, the result can be generalized to $\delta\in (0,1)$\cite{borthwick2007spectral}. Furthermore, Huber and McKean proved that there is a correspondence between the Laplacian spectrum $\lambda_i$ of $X$ and the geodesic length spectrum $\ell_i([T_i])$ where $[T]$ is the equivalent class of $T$ for all conjugate element in $\Gamma$\cite{huber1959analytischen,mckean1972selberg,MR473167}. Lastly, there is a correspondence between the equivalence classes $[T_i]$ and the geodesic length of $[T_i]$\cite{dal2010geodesic}.

The main theorem that is showed in the next two sections is the following
\begin{theorem}[Main Theorem]
	Let a partition function of an Ising Model be given. Then, a corresponding partition function of an ensemble of equivalent classes of hyperbolic surfaces can be uniquely constructed. The converse is also true if the given hyperbolic surfaces are all generated by Schottky groups with half-spaces located on the places that can be described by the Ising model on the boundary.
	
	Furthermore, let $\left\lbrace\partial X_\alpha\right\rbrace_{\alpha=1}^\infty$ be the index set of all possible configurations of the Ising model on the boundary, and $\left\lbrace X_\alpha\right\rbrace_{\alpha=1}^\infty$ be the index set of all possible corresponding equivalent classes of hyperbolic surfaces. Then, in both directions, there exists an invariant of probability measures:
	$$
	\int_{ \partial X_\alpha} d\mu_{\partial X}=\int_{ X_\alpha } d\mu_{X}.
	$$
\end{theorem}

The arguments of the proof in the following two sections are written for two-dimensional hyperbolic space and its boundary.

\section{From Ising Models to Hyperbolic Surfaces}
The goal of this section is to enumerate all classical hyperbolic surfaces $\Gamma_\alpha\setminus\mathbb{D}$, $\alpha\in\mathbb{N}$, where $\Gamma_\alpha$ is determined by the Ising model on the boundary of $\mathbb{D}$ and $\Gamma_\alpha$ is an equivalent class of Schottcky groups under conjugations such that elements of this equivalent class all have the same configuration of half-spaces, i.e. these conjugations are determined by the elements in $\text{PSU}(1,1)$ that can preserve the given configuration on $\partial\mathbb{D}$.

On the boundary $\partial\mathbb{D}$, let $N\in\mathbb{N}$ be given. Assume that $\mathbb{D}$ is embedded in $\mathbb{C}$, thus $\partial\mathbb{D}$ is in the Euclidean space. Uniformly divide $\partial\mathbb{D}$ into $N$ cells such that all of them have the same size under Lebesgue measure, and assign each cell a spin $\sigma_i$, where $\sigma_i\in\left\lbrace -1, +1\right\rbrace$.

Denote each cell by $D_i$. Systematically pick points on $\partial\mathbb{D}$ using cells $D_i$ such that $+1$ and $-1$ are systematically assigned to each $D_i$ to obtain all possible configurations for constructing an Ising model. For example, if $N=3$, the values in $D_1, D_2,$ and $D_3$ are
$$(1,1,1),(1,1,-1),$$
$$(1,-1,1),(1,-1,-1),$$
$$(-1,-1,-1),(-1,1,1),$$
$$(-1,1,-1), \text{ and }(-1,-1,1).$$ For each $D_i$ with $\sigma_i=1$, assume there is a geodesic connects the two boundary points of $D_i$ on $\partial \mathbb{D}$. In 3d, these boundary points form a circle, and the arbitrary union of all geodesics form a surface of a partial sphere.

Each configuration of these half-spaces $D_i$ is also an equivalent class under all conformal isometries in $\text{PSU}(1,1)$ that can preserve the configuration on the unit circle $\partial\mathbb{D}$. For some symmetric configurations, these mappings could include not only rotations but reflections, otherwise they are combinations of M\"{o}bius mappings, i.e. combinations of reflections. The result is for each configuration, the cardinality of the set of all these conjugations is the same as the cardinality of the set of all conjugations applied to a representative in $[\Gamma_\alpha]$, since for each conjugation that is applied to the boundary $\partial\mathbb{D}$ is also applied to the interior $\mathbb{D}$, and the conjugation of the interior is determined by the conjugation applied to the boundary.

Referring back to the $N=3$ example, the limit set is empty when configured as $(-1,-1,-1)$. In contrast, the limit set of the configuration $(1,1,1)$ is covered by three half-spaces that intercept each other's boundary on $\partial\mathbb{D}$ as illustrated in \cite{mcmullen1998hausdorff} rather than equal to $\partial\mathbb{D}$. Each group generator is a reflection if $N$ is odd; McMullen's 3-pant Schottky group is one well-known example\cite{mcmullen1998hausdorff}.

In Dal'Bo's book \cite{dal2010geodesic} and the author's MA thesis\cite{chuang2022hausdorff}, there are further instances of creating Schottky groups using half-spaces with different $N$.

Then, the appropriate Riemannian metric of the hyperbolic surface may then be computed using Poincar\'{e}'s theorem, and the metric can be substituted into Einstein's equation after the $N$ group generators have been established using the approach illustrated in \cite{dal2010geodesic,chuang2022hausdorff}.

As a result, it is possible to derive the integral of the Einstein-Hilbert action, allowing for a numerical calculation of the path integral technique to be systematically examined in this setup. Furthermore, matter fields may need to be added (particularly if instead of using partition function but path integral for an application in quantum field theory version) so that both sides can have the same probability distribution as $N$ goes to infinity in order to ensure that the probability of a particular configuration of the Ising system on the boundary is identical to the outcome obtained from the configuration of algebraic curves. If path integrals are used on both sides, each path will be measured, and the value will be referred to as the amplitude by convention. The following assumes that both sides are written in partition functions because proving the existence of the invariant is the goal of this note. In principle, this derivation can be generalized to a path integral version with a careful definition on measures and integrands, particularly on the side of the algebraic curves $\Gamma_\alpha\setminus\mathbb{D}$.

In other words, by Poincar\'{e}'s Uniformization theorem and Patterson-Sullivan's theorem, the geometry of a hyperbolic surface (a kind of Riemann surfaces and algebraic curves) $\Gamma_\alpha\setminus\mathbb{D}$ is determined by the limit set $L(\Gamma_\alpha)\subset \partial\mathbb{D}$, and the limit set is determined by the given configuration of Ising model, i.e. the existence of the hyperbolic surface is determined by the existence of the corresponding configuration of Ising model, so the probability (or amplitude, if path integral was used) for each of them to occur should be identical.

Thus, we can define measures on both sides to be $\sigma$-finite such that for each $\alpha\in\mathbb{N}$ we have
$$
\int_{ \partial X_\alpha} d\mu_{\partial X}=\int_{ X_\alpha } d\mu_{X}
$$
where $\left\lbrace\partial X_\alpha\right\rbrace_{\alpha=1}^\infty$ is the index set of all possible configurations on the boundary, and $\left\lbrace X_\alpha\right\rbrace_{\alpha=1}^\infty$ is the index set of all possible equivalent classes of hyperbolic surfaces $\Gamma_\alpha\setminus \mathbb{D}$. Since there is a one to one correspondence between $\partial X_\alpha$ and $X_\alpha$, we have
$$Z_{\text{Ising}}=\int_{\bigcup\limits_{\alpha=1}^\infty \left\lbrace\partial X_\alpha\right\rbrace} d\mu_{\partial X}=\int_{\bigcup\limits_{\alpha=1}^\infty \left\lbrace X_\alpha \right\rbrace} d\mu_{X}=Z_{\text{Hyperbolic Surfaces}}.$$

\section{From Hyperbolic Surfaces to Ising Models}
The goal of this section is to compute the partition function of an Ising model using a given collection of hyperbolic surfaces that are constructed by Schottky groups. Because by the correspondence theorem\cite{sullivan1979density, bishop1997hausdorff} the set-up in the previous section can be used conversely.

Hence, conversely, with the same notations as defined in the previous section, given an ensemble of compact hyperbolic surfaces with half-spaces located on the places that can be described by Ising model on the boundary as defined in the previous section, for each configuration one can reduce it to fundamental polygon using the method in the proof of classification theorem, and based on Poincar\'{e}'s theorem, we can reconstruct the Schottky group $\Gamma_\alpha$ which is an equivalent class under conjugation. In this direction, the conjugation of the interior is still determined by the conjugation applied to the boundary. Since the hyperbolic surfaces that are constructed by Schottky groups are given, so for each Schottky group we could find their half-spaces determined by group generators\cite{beardon2012geometry,dal2010geodesic}.

Then for each given $N$, since for simplicity, the author assumes all half-spaces are located on the places that can be described by Ising model on the boundary, the only cases is that for a cell of Ising model, either there is a half-space coincide to it, or the interception of all half-spaces to that cell is an empty set. Hence, in the former case the corresponding spin $\sigma_i$ of that cell is assigned to $+1$, otherwise $\sigma_i:=-1$. In 1-dimension, each cell is a segment of the circumstance of the unit circle, and all cells are evenly distributed on the circumstance of the unit circle $\partial\mathbb{D}$.

Next, let $H(\sigma)=-\sum\limits_{\left\langle i,j \right\rangle}J_{ij}\sigma_i\sigma_j-\mu\sum\limits_{j}h_j\sigma_j$ be the Hamiltonian of an Ising model where $J_ij$ is the coupling constant of the interaction, $h_j$ is the field strength of an external field, and historically $\mu$ is called magnetic moment. In our set-up it is a generic parameter. The goal is to tune these parameters such that both systems, i.e. the ensembles of Ising model and hyperbolic surfaces, have an identical probability distribution, i.e. the result is the invariant $\int_{ \partial X_\alpha} d\mu_{\partial X}=\int_{ X_\alpha } d\mu_{X}$. This can be done by the Poincar\'{e}'s Uniformization theorem and Patterson-Sullivan's theorem. Since, conversely, the limit set of a Schottky group determined the corresponding configuration of the Ising model. Hence, the probability for each side to occur should be identical. Hence, we have this invariant. The result of this direction is that computing the probability of certain hyperbolic manifolds gives the probability of certain configurations in the Ising model defined on the boundary.

\begin{corollary}
	The restriction on the assumption of the locations of half-spaces of each Schottky group can be relaxed.
\end{corollary}
The invariant can still be valid as long as the ensembles of the Ising model and equivalent classes of hyperbolic surfaces have equal cardinality and are well-defined.

For instance, one may apply a coarse-grained algorithm to half-spaces that are adjacent to each other and allow assigning spins to half-spaces with different sizes.

\section{Acknowledgment}
The first draft of this note was written as a combination of three final projects while attending Prof. Pisin Chen's Cosmological Physics and an independent/advanced study course and Prof. Ning-Ning Pang's Statistical Mechanics at National Taiwan University. The original purpose was twofold: one, to mathematically explain and, if feasible, prove the AdS/CFT correspondence; secondly, to show the connection between gravity and number theory based on statistical field theory. AdS/CFT is stronger than the result in this note since it requires supersymmetry. The author appreciates the opportunity offered by Prof. Chen to study within his research groups at NTU. Lastly, many thanks to everyone who has helped the author better comprehend this issue over the last 10 years.

\bibliographystyle{amsplain} 
\bibliography{refs} 

\end{document}